\documentclass[twocolumn,showpacs,amsmath,amssymb]{revtex4}
\usepackage[dvips]{graphicx}
\usepackage{bm}

\begin{document}
\title{Nonequilibrium Kondo Effect in a Quantum Dot Coupled to Ferromagnetic Leads}
\author{
Yasuhiro Utsumi$^{1,2}$,
Jan Martinek$^{1,3,4}$,
Gerd Sch\"on$^1$,
Hiroshi Imamura$^5$,
and
Sadamichi Maekawa$^3$
}
\affiliation{
$^1$ Institut f\"ur Theoretische Festk\"{o}perphysik, Universit\"{a}t Karlsruhe, D-76128 Karlsruhe, Germany \\
$^2$ Max-Planck-Institut f\"ur Mikrostrukturphysik, Weinberg 2, D-06120 Halle (Saale), Germany \\
$^3$ Institute~for~Materials~Research,~Tohoku~University,~Sendai~980-8577,
Japan \\
$^4$ Institute of Molecular Physics, Polish Academy of Sciences, 60-179 Pozna\'n, Poland \\
$^5$ Graduate~School~of~Information~Sciences,~Tohoku~University,~Sendai~980-8577,
Japan
}

\begin{abstract}
We study the Kondo effect in the electron transport through a quantum dot
coupled to ferromagnetic leads, using a real-time diagrammatic technique which provides a systematic description of the nonequilibrium dynamics of a system with strong local electron correlations. We evaluate the theory in an extension of the `resonant tunneling approximation', introduced earlier, by introducing the self-energy of the off-diagonal component of the reduced propagator in spin space. In this way we develop a charge and spin conserving approximation that accounts not only for Kondo correlations but also for the spin splitting and spin accumulation out of equilibrium. We show that the Kondo resonances, split by the applied bias voltage, may be spin polarized. A left-right asymmetry in the coupling strength and/or spin polarization of the electrodes significantly affects both the spin accumulation and the weight of the split Kondo resonances out of equilibrium. The effects are observable in the nonlinear differential conductance. We also discuss the influence of decoherence on the Kondo resonance in the frame of the real-time formulation.
\end{abstract}

\date{\today}
\maketitle

\newcommand{\rd}{d}
\newcommand{\ri}{i}
\newcommand{\mat}[1]{\mbox{\boldmath$#1$}}
\newcommand{\varepsilou}{\omega}
\newcommand{\etb}{0^+}

\section{introduction}
\label{sec:introduction}



The continuing experimental progress with mesoscopic electronic devices 
has stimulated anew the interest in
fundamental quantum mechanical questions~\cite{ImryBook}.
A good example is the Kondo effect~\cite{Hewson} in electron transport 
through quantum dots (QD). 
The effect, which nowadays is well established experimentally~\cite{GoldhaberGordon}, 
has remained a challenging problem for 
nonequilibrium quantum transport theory
~\cite{Hershfield0,Meir1,Meir2,Konig1}.
After early experiments, which concentrated on semiconductor QDs,
the Kondo effect was observed in single-atom~\cite{Park}
and single-molecule transistors~\cite{Liang}, 
as well as in carbon nanotube QDs~\cite{Nygard}, 
all of which were coupled to metallic leads. 
More recently, the Kondo effect was demonstrated in QDs
coupled to ferromagnetic leads~\cite{Pasupathy,Nygard1}. 
In this article, we will discuss the nonequilibrium Kondo 
effect in such systems as an example of the 
nontrivial collective many-body physics in magnetic nanostructures. 

\begin{figure}[t]
\includegraphics[width=0.70\columnwidth]{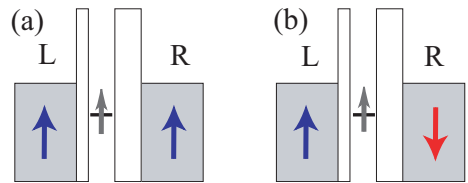}
\caption{ A single level quantum dot coupled to two ferromagnetic
leads for (a) parallel and (b) antiparallel alignments of lead
magnetizations. 
We allow for a left-right asymmetry in tunnel
barriers and/or spin polarization factors. } \label{fig:QD}
\end{figure}

Research on electronic properties in magnetic systems 
has evolved into an active field, \lq\lq spintronics"~\cite{Maekawabook}. 
In spintronics devices 
magnetic properties control transport properties 
via the electron spin degree of freedom.
One example is the tunnel magnetoresistance (TMR)
in ferromagnetic tunnel junctions~\cite{Miyazaki},
which relies on switching the magnetization 
of the electrodes between parallel (P) and antiparallel (AP) orientations by an applied magnetic field.
In magnetic multilayers, 
the intrinsic exchange coupling between layers~\cite{Bruno1}
influences the magnetic structure 
and consequently the transport properties~\cite{Shinjo}.
%
%
The Kondo effect in transport through a QD coupled to 
ferromagnetic leads (Fig.~\ref{fig:QD}) 
provides a prime example of many-body effects in spintronics devices. 
The theoretical model, a local spin coupled to two ferromagnetic
leads, was originally introduced
to explain the decrease of the TMR with increasing temperature~\cite{Inoue}. 
The model was revived recently, and led 
 to a series of publications
~\cite{Sergueev,Martinek1,Dong,PZhang,Lopez,Martinek2,MSChoi}.
In the early stage, due to different approximations used, the question remained controversial whether the Kondo resonance splits in the presence 
of spin-polarized leads~\cite{Martinek1,Dong} or not~\cite{PZhang,Lopez}. The splitting was 
confirmed eventually by numerical-renormalization-group (NRG) techniques~\cite{Martinek2,MSChoi}.

The origin of the splitting of the Kondo resonances lies
an exchange coupling induced by spin-dependent quantum charge fluctuations. 
Let us consider a single-level, originally spin-degenerate QD with
energy $\epsilon_0\! < \! 0$ (measured relative to
the Fermi level of the lead electrons) and  strong Coulomb
interaction, $U \rightarrow \infty$. Then, double occupancy is
suppressed and the QD behaves as a magnetic impurity. But there 
still exist quantum charge fluctuations due to the tunneling 
between the QD and the leads.
An electron with majority-spin in the QD can tunnel
between the QD and the leads easier than an electron with
minority-spin, thus gaining a larger kinetic energy. 
As a result, a ferromagnetic exchange interaction 
between the QD spin and the lead magnetization is induced. 
It lifts the spin degeneracy of the QD level by $\Delta \epsilon$.
Such a spin splitting quenches spin-flip scattering processes for
low energies, below $|\Delta \epsilon|$, and suppresses the Kondo correlations. It also leads to a splitting of the Kondo resonance.
The qualitative two-stage scaling theory presented in Ref.~\cite{Martinek1}
indicates
that the high-energy part of quantum charge fluctuations, between
$|\epsilon_0|$ and a cutoff energy $D$ (of the order of
the itinerant electron band width)
is responsible for a spin splitting given by
\begin{eqnarray}
\Delta \epsilon \sim -P \frac{\Gamma}{\pi} \ln
\frac{D}{|\epsilon_0|}, \;\;\;\; \Gamma \! \equiv \! (\Gamma_{
\uparrow}+\Gamma_{ \downarrow})/2 \, . \label{eqn:spliting}
\end{eqnarray}
Here we assumed the coupling strength $\Gamma_{ \sigma}$ between 
the QD level and the leads to be spin dependent. 
The splitting of the level results in a corresponding 
spin splitting of the Kondo peak in the spectral density. 
Remarkably, it is possible to compensate the spin splitting 
$\Delta \epsilon$ by applying an external magnetic field, 
and thus to recover the full Kondo effect~\cite{Martinek1}.

In spite of the interest in the problem, a {\sl systematic}
approximation which can account for the spin splitting of the Kondo effect 
in nonequilibrium transport problems has not yet been presented.\\
- The NRG is a powerful tool for the investigation of
the Kondo problem, but is limited to the equilibrium state. 
The validity of extensions to the nonequilibrium 
state~\cite{Lopez,Dong,Martinek1} remains to be proven.\\
- The slave-boson mean-field approximation is correct in the
limit of large degeneracy and able to capture the physics of the
Kondo singlet at zero temperature~\cite{Coleman}. 
However, it does not account for the spin splitting since 
the description of quantum charge fluctuations
is beyond mean-field theory.\\
- The non-crossing approximation (NCA), a systematic conserving
approximation~\cite{Kuramoto,Bickers}, has
been successfully adopted to the nonequilibrium state~\cite{Meir1,Meir2}. 
NCA accounts for spin-dependent quantum charge fluctuations and thus 
for the spin splitting due to ferromagnetic leads. 
But, it also generates 
an additional spurious peak at the Fermi level 
even in equilibrium~\cite{Dong}
as well as 
a spurious peak in the presence of magnetic field~\cite{Meir1}.
Such a shortcoming is dangerous in the 
context of transport, which is sensitive to the
electron states near the Fermi energy.
The spurious peak can be made to disappear, 
if vertex corrections are properly accounted for~\cite{Kroha}. 
However, the generalization of NCA described in Ref.~\cite{Kroha}
does not appear tractable in nonequilibrium situations.
\\
- The equation-of-motion (EOM) approach, 
which was adopted in Ref.~\cite{Martinek1},  
is able to capture Kondo correlations as well as -- in principle -- the spin splitting. However, within the standard
decoupling scheme, the spin
splitting is lost~\cite{Sergueev,PZhang}, and it
requires introducing an additional self-consistent equation
to determine the renormalized QD-level energy~\cite{Martinek1}.

In this situation, a systematic approximation, free of the mentioned
drawbacks in nonequilibrium, is needed for further progress in the
exploration of the Kondo effect in spintronics devices.
In this paper, we formulate the problem using a real-time
diagrammatic technique, which is a systematic method suitable 
to describe the nonequilibrium time evolution of a system 
with strong local electron correlations~\cite{Schoeller_Schon,Konig1}. 
We evaluate the theory, accounting for Kondo correlations and 
spin-dependent quantum charge fluctuations,
in a conserving approximation
making use of an extension of the resonant tunneling
approximation (RTA)~\cite{Konig1}. The RTA 
was shown to produce qualitatively
reasonable results in the presence of a magnetic field
at not too low temperatures, sufficient to allow for a comparison 
with experiment~\cite{Ralph}. 
This property motivated us to apply the RTA for the present problem.

The outline of this paper is as follows.
In Sec.~\ref{sec:cal}, we introduce the model Hamiltonian
and shortly review the real-time diagrammatic technique
as well as the resonant tunneling approximation.
We introduce an extension of the RTA, which is needed to
account for the spin splitting.
In Sec.~\ref{sec:result} we present numerical results.
We will show the splitting of the zero-bias anomaly
for various temperatures (Sec.~\ref{sec:splitting}).
We provide a comprehensive discussion of
the restoration of the Kondo resonance by an applied
magnetic field (Sec.~\ref{sec:restoration})
and the nonequilibrium Kondo effect (Sec.~\ref{sec:nonlinear}).
In Sec.~\ref{sec:asym}, we will discuss the effect of
asymmetries in system parameters and their consequences for nonequilibrium states.
In Sec.~\ref{sec:decoherence}, we will discuss decoherence effects
on the Kondo resonance out of equilibrium within the framework of 
the real-time diagrammatic technique.
Sec.~\ref{sec:exp} is devoted to a discussion 
of the relation between our results and 
recent experiments~\cite{Pasupathy,Nygard}. 
We summarize in Sec.~\ref{sec:sum}.

\section{formulation}
\label{sec:cal}

\subsection{Model Hamiltonian}

Our model Hamiltonian consists of several parts describing a
single-level QD, $H_{\rm D}$, the left (and right) reservoir,
$H_{L(R)}$, and tunneling, $H_T$,
\begin{equation}
H
=
H_{\rm D}+\sum_{r=L,R} H_r+H_{\rm T}.
\end{equation}
%
The Hamiltonians for the isolated QD and the leads are
\begin{eqnarray}
H_{\rm D} \!\! &=& \!\! \sum_{\sigma} \epsilon_\sigma
d_\sigma^\dagger d_\sigma + U d_\uparrow^\dagger d_\uparrow
d_\downarrow^\dagger d_\downarrow \, ,
\\
H_r
\!\! &=& \!\!
\sum_{r k \sigma}
\varepsilon_{r k \sigma} \, a^{\dagger}_{r k \sigma} a_{r k \sigma} \, ,
\end{eqnarray}
respectively
(we use unit where $\hbar \!=\! k_{\rm B} \!=\! 1$). 
Here, $d_\sigma$ is the annihilation operator of electrons
with spin-$\sigma$ ($\sigma \! = \uparrow, \downarrow$)
in the QD, and $U$ denotes the onsite Coulomb
interaction.
The energy of the single QD level with spin 
$\uparrow$/$\downarrow$ is
$\epsilon_{\uparrow/\downarrow} \! = \! \epsilon_0 \pm E_{\rm
Z}/2+a_c eV/2$. The first term, $\epsilon_0$, may be tuned
by a gate voltage. The Zeeman energy is given by $E_{\rm Z}=2 g
\mu_{\rm B} B$, where 
the external magnetic field, $B$, is assumed to act (only) 
on the spin in the QD. 
The applied bias voltage
between the left and right leads, $V$,
may shift the QD level by $a_c eV/2$~\cite{Konig1}.
Here $a_c$ is an asymmetry factor due to different capacitive couplings of the QD
to the left and right leads.
The operator $a_{r k \sigma}$ annihilates 
an electron in the lead $r \! = \! L,R$ with wave number $k$ 
and spin $\sigma$. We assumed that the
ferromagnetism in the leads can be treated within mean-field
approximation, i.e., it is accounted for by a
spin-dependent dispersion $\varepsilon_{r k \sigma}$ and, 
hence, spin-dependent density of states (DOS) $\nu_{r \sigma}
(\varepsilou) \!=\! \sum_k
\delta(\varepsilou-\varepsilon_{r k \sigma})$. 

The tunneling Hamiltonian, finally, is given by
\begin{equation}
H_{\rm T}
=
\sum_{r k \sigma}
V_{r k} \, d_\sigma^\dagger
a_{r k \sigma} +{\rm H.c.}
\label{eqn:tunnelinghamiltonian}
\end{equation}
We assumed that the tunneling amplitude $V_{r k}$
is independent of spin.
Furthermore, for simplicity, we will ignore a $k$ dependence of the
tunneling amplitude $V_{r k} \!=\! V_r$.

The important parameter of this model is 
the spin-dependent coupling strength to the ferromagnetic lead $r$,
\begin{equation}
\Gamma_{r \sigma}(\varepsilou) \! \equiv \! 2 \pi |V_r|^2 \nu_{r
\sigma}(\varepsilou) \; . 
\label{eqn:coupling}
\end{equation}
The spin dependence is 
the main generalization required for the present problem as compared to 
that of a QD coupled to normal metal leads~\cite{Konig1}.
The parameter
$\Gamma_\sigma$, which appeared in Eq.~(\ref{eqn:spliting}), 
is given by
$\Gamma_\sigma \equiv \sum_r \Gamma_{r \sigma}$, 
where $\Gamma_{r \sigma} \! \equiv \! \Gamma_{r \sigma}(0)$ is 
the value at the Fermi energy. 
The spin-dependent coupling strengths are 
related to the spin polarization of the leads by
\begin{eqnarray}
P_r = 
(\Gamma_{r \uparrow}-\Gamma_{r \downarrow})/
(\Gamma_{r \uparrow}+\Gamma_{r \downarrow}) \; .
\end{eqnarray}
%
In this paper, we will analyze two configurations of the lead
magnetizations, the parallel (P) and antiparallel (AP) alignments.
We always take a positive value for $P_L$. The P (AP) alignment then
corresponds to positive (negative) values of $P_R$.

\subsection{Real-time diagrammatic technique}

For infinitely strong Coulomb interaction, $U\!=\!\infty$, 
double occupancy of the QD level is suppressed, and the Hilbert 
space is restricted to the
empty state $|0 \rangle$ and two singly occupied states $|\sigma
\rangle$. 
In the presence of strong interaction effect, the application of
Wick's theorem is prohibited, and the standard approach for the
calculation of nonequilibrium currents through a tunnel junction
~\cite{Caroli_1} is not suitable. We, therefore, proceed using
the \lq real-time diagrammatic technique' developed by Schoeller 
{\it et al.}~\cite{Schoeller_Schon,Konig1}.
Apart from providing a transparent classification of various processes  
it allows constructing a conserving approximation out of
equilibrium.
In the following, we will briefly outline the technique.
It is closely related the Keldysh formalism and the Feynmann-Vernon approach
(see for example Ref.~\cite{Weiss}).
The technique enables one to perform a systematic diagrammatic 
expansion of the reduced density matrix 
in terms of $H_{\rm T}$ in the real-time domain.

The central quantity is the reduced density matrix, 
which is derived from the total density
matrix by tracing out the lead electron degrees 
of freedom,
\begin{eqnarray}
P^\chi_{\chi'}(t)
=
{\rm Tr}_L {\rm Tr}_R 
{\rm Tr}_D
\left[ \rho^L_0  \rho^R_0 \rho^{\rm D}_0 \,
T_C S_C 
X_{\chi' \, \chi} (t)_I
\right],
\label{eqn:prob}
\\
X_{\chi' \, \chi} (t)_I
\equiv
{\rm e}^{ \ri H_0 t}
|\chi' \rangle \langle \chi|
{\rm e}^{-\ri H_0 t}          
\, , \;
(\chi,\chi' \! = \! 0,\uparrow,\downarrow).
\label{eqn:Hubberd}
\end{eqnarray}
The combination of a trace over dot states, ${\rm Tr}_D$, and 
a projection operator, $X_{\chi' \, \chi}$, 
selects the $(\chi,\chi')$-component of the reduced density matrix 
$P^\chi_{\chi'}$. 
The subscript $I$ denotes the interaction picture 
with respect to the unperturbed Hamiltonian
$H_0 \! \equiv \! H_L  \! +  \! H_R \! + \! H_{\rm D}$. 
The contour ordering operator $T_C$ is defined on the closed
time-path $C$ starting from the initial time, $t_i$, to time $t$ on
the upper branch of the Keldysh contour and returning to $t_i$ on
the lower branch. The Keldysh $S$-matrix is
defined along this contour by
\begin{equation}
S_C \equiv \exp \left( -\ri \! \int_C \! \rd t
H_{\rm T}(t)_I \right).
\end{equation}
Initially, at time $t_i$, the density matrix is assumed to be factorized into
a product of equilibrium density matrices of the leads,
$\rho^r_0$, and the density matrix of the QD, $\rho^{\rm D}_0$.
The former are characterized by the temperature $T$ and chemical
potential $\mu_r$ ($\mu_L \! = \! - \mu_R \! = \! eV/2$) as,
\begin{equation}
\rho^r_0 \equiv {\rm e}^{-\beta (H_r-\sum_{k \sigma} \mu_r
a^{\dagger}_{r k \sigma} a_{r k \sigma})}/Z^r_0 \; ,
\end{equation}
where $\beta \! \equiv \! T^{-1}$. 
The normalization factor $Z^r_0$ ensures that
${\rm Tr}_r \rho^r_0\!=\!1$.

Equation~(\ref{eqn:prob}) can be evaluated in a perturbative
expansion in terms of $H_{\rm T}$. 
Under the conditions that the initial density matrix of the 
QD is diagonal, $\langle \chi'|\rho^{\rm D}_0|\chi \rangle \! = \!
P^0_\chi \, \delta_{\chi',\chi}$, 
and that the Hamiltonian conserves the spin, 
the time evolution reduces in the limit of $t_i \rightarrow -\infty$
to a master equation: 
\begin{eqnarray}
\sum_{\chi =0,\uparrow,\downarrow} P^{\rm st}_{\chi} \,
\Sigma_{\chi,\chi'} \! = \! 0 \; ,
 \label{eqn:master}
\end{eqnarray}
for the (stationary) probability $P^{\rm st}_\chi$ which is the diagonal 
($\chi,\chi$)-component of the stationary density matrix. 
The self-energy of the master
equation $\Sigma_{\chi,\chi'}$ is related to transition
probabilities from a state $|\chi \rangle$ to a state $|\chi'
\rangle$.
The main task is to determine an appropriate self-energy of the
master equation in the frame of the diagrammatic expansion.

\begin{figure}[t]
\includegraphics[width=0.90\columnwidth]{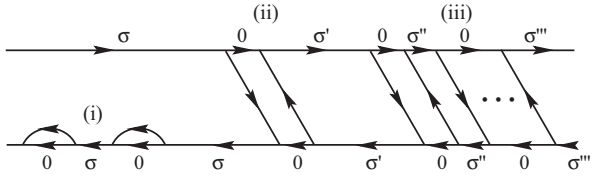}
\caption{ Diagrams for various tunneling processes: (i) quantum
charge fluctuations, (ii) the co-tunneling process and (iii) the
resonant tunneling process. 
The horizontal lines on the Keldysh contour
represent the free propagators of the quantum dot. Remaining directed lines are tunneling lines.} \label{fig:diagram}
\end{figure}

Figure~\ref{fig:diagram} shows examples of diagrams. Basic
building blocks are the free evolution (horizontal lines
on the Keldysh contour) and the tunneling processes 
(remaining directed lines) of electrons with 
spin $\sigma$ into or out of the QD 
to the leads $r$ ~\cite{Konig1}. 
In Fourier space the latter carry a factor obtained 
by Fermi's golden rule,
\begin{equation}
\gamma^\pm_{r \sigma}(\varepsilou)
=
\Gamma_{r \sigma}(\varepsilou)
f^\pm(\varepsilou-\mu_r)/(2 \pi),
\label{eqn:tunnelingline}
\end{equation}
where $f^\pm(\varepsilou) \! = \! 1/({\rm e}^{\pm \beta
\varepsilou} \! + \! 1)$ is the electron (hole)
distribution function. 
In oder to regularize the expressions 
we impose a Lorentzian cutoff replacing the coupling strengths
of Eq.~(\ref{eqn:coupling}) by 
$\Gamma_{r \sigma} \rho_{\rm c}(\varepsilou)$ 
where
$
\rho_{\rm c}(\varepsilou) \! = \! D^2/(\varepsilou^2 \! + \! D^2).
$
The effective bandwidth $D$ was already introduced in
Eq.~(\ref{eqn:spliting}). 

In Fig.~\ref{fig:diagram}, each sector of the diagram corresponds
to a physical tunneling process. The \lq bubbles' in sector (i)
represent quantum charge fluctuations, where an electron with spin $\sigma$
virtually fluctuates between the QD and the leads. A single line 
connecting the upper and lower horizontal lines (not shown) 
describes a sequential tunneling process. Two overlapping
parallel lines shown in sector (ii)
represent a co-tunneling process, where in one coherent transition 
an electron with spin
$\sigma$ leaves the QD and an electron with spin
$\sigma'$ fills the empty level. A \lq train' of
overlapping lines  [sector (iii)] represents a process
with electrons going back and forth many times between the QD and the leads.

\subsection{Extension of resonant tunneling approximation}

In order to account for the logarithmic behavior of Kondo correlations, we
have to sum up a suitable class of diagrams of the expansion up to
infinite order. The starting point of our approximation is the class of diagrams denoted as \lq resonant tunneling approximation' (RTA)~\cite{Konig1}. 
In the diagrammatic language, 
RTA includes all diagrams where any
auxiliary vertical line cuts {\it at most two tunneling lines}
[An example is presented in Fig.~\ref{fig:diagram}(iii)].
At not too low temperatures RTA was shown to provide 
a qualitatively reasonable approximation for the Kondo 
correlations with analytical expressions for the nonlinear differential conductance for arbitrary level degeneracy.
It predicts also qualitatively reasonable behavior 
in the presence of Zeeman splitting. However, the original RTA developed in
Ref.~\cite{Konig1} does not describe the spin splitting 
since, as we will discuss below, it does not take into account the
subclass of diagrams which accounts for spin-dependent quantum 
charge fluctuations.

\begin{figure}[t]
\includegraphics[width=1.0\columnwidth]{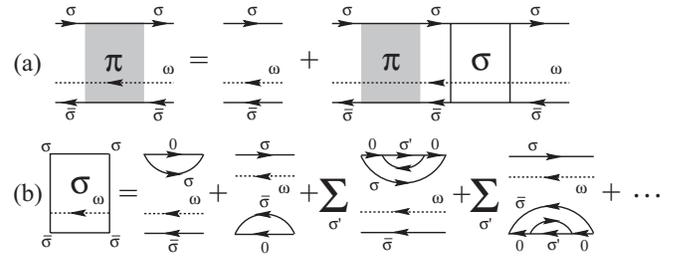}
\caption{ (a) Dyson equation for
$\pi^{\sigma,\sigma}_{\bar{\sigma},\bar{\sigma}}$ and (b) the
diagrammatic expansion of the irreducible self-energy
$\sigma^{\sigma,\sigma}_{\bar{\sigma},\bar{\sigma}}$.
Directed dotted lines carry the energy $\varepsilou$. }
\label{fig:sfgf}
\end{figure}

To find the origin of the shortcoming, let us examine the QD spin
dynamics. 
In general, in order to describe the superposition of spin states 
$|\! \uparrow \rangle$ and $|\! \downarrow \rangle$, 
it is necessary to consider 
also off-diagonal components of the density matrix in spin space, e.g., 
$P^\sigma_{\bar{\sigma}}$, where $\bar{\sigma}$ is the opposite spin of
$\sigma$. 
The time evolution of $P^\sigma_{\bar{\sigma}}$ between
time $t_i$ and $t$ ($t>t_i$) is described by the  reduced
propagator 
\begin{equation}
\pi^{\sigma,\sigma}_{\bar{\sigma},\bar{\sigma}}(t,t_i)
\! = \!
-\ri
\langle \bar{\sigma}|
{\rm Tr}_L {\rm Tr}_R \left[ \rho^L_0  \rho^R_0 \, T_{C} \, S_{C}
X_{\bar{\sigma} \, \sigma} (t)_I \right] 
|\sigma \rangle.
 \label{eqn:rp}
\end{equation}
Since the tunneling Hamiltonian Eq.~(\ref{eqn:tunnelinghamiltonian})
conserves the spin, spin is conserved at each vertex. E.g., the spin
indices of an incoming/outgoing tunneling line 
and an outgoing/incoming free propagator 
for each vertex are the same. Therefore,
after Fourier transformation, 
Eq.~(\ref{eqn:rp}) can be expressed as a Dyson equation,
\begin{equation}
\pi^{\sigma,\sigma}_{\bar{\sigma},\bar{\sigma}}(\varepsilou)
=
1/
(\varepsilou+\ri \etb
-\epsilon_\sigma+\epsilon_{\bar{\sigma}}
-\sigma^{\sigma,\sigma}_{\bar{\sigma},\bar{\sigma}}(\varepsilou)
),
\label{eqn:sppr}
\end{equation}
where
$\sigma^{\sigma,\sigma}_{\bar{\sigma},\bar{\sigma}}$ denotes an
irreducible self-energy 
and $\etb$ is a positive infinitesimal number. 
The diagrammatic representation of Eq.~(\ref{eqn:sppr}) is
given in Fig.~\ref{fig:sfgf}(a). 
Directed dotted lines carrying the energy $\varepsilou$
correspond to
the factor $\exp ( \ri \varepsilou t)$ of the Fourier transformation.
The diagrams for $\sigma^{\sigma,\sigma}_{\bar{\sigma},\bar{\sigma}}$
are shown in Fig.~\ref{fig:sfgf}(b),
in which any auxiliary vertical line cuts
at least one tunneling line.

Now, we observe that within RTA in any diagram for the 
self-energy of the master equation Eq.~(\ref{eqn:master}) 
a sector, where the reduced propagator
$\pi^{\sigma,\sigma}_{\bar{\sigma},\bar{\sigma}}$
appears, 
is always accompanied by two tunneling lines,
a left-going tunneling line with spin $\sigma$
and
a right-going tunneling line with spin $\bar{\sigma}$,
which are disconnected from
$\pi^{\sigma,\sigma}_{\bar{\sigma},\bar{\sigma}}$
itself.
It is a consequence of the requirement that 
any auxiliary vertical line must cut the same number
of left- and right-going tunneling lines or free propagators
with the same spin indices. 
This in turn follows from the fact that the spin
is conserved at each vertex, and that the initial QD density
matrix is assumed to be diagonal.
As a result, {\it the irreducible self-energy
$\sigma^{\sigma,\sigma}_{\bar{\sigma},\bar{\sigma}}$ 
does not appear in RTA}~\cite{Konig1}, 
because in its presence, 
any auxiliary vertical line laying on a sector of 
$\sigma^{\sigma,\sigma}_{\bar{\sigma},\bar{\sigma}}$ 
would cut {\it at least three tunneling lines}. 
RTA can account for the Zeeman
splitting of the Kondo resonance, which is already included in a
denominator of the bare reduced propagator
$\pi^{\sigma,\sigma}_{\bar{\sigma},\bar{\sigma}}$. 
However, RTA omits further information on the spin dynamics. 
As we will discuss in the following, the further information, 
such as the spin splitting $\Delta \epsilon$ 
caused by spin-dependent quantum charge fluctuations, 
is included in $\sigma^{\sigma,\sigma}_{\bar{\sigma},\bar{\sigma}}$
beyond the original RTA.

In order to obtain an expression for
$\sigma^{\sigma,\sigma}_{\bar{\sigma},\bar{\sigma}}$ 
we proceed in a perturbative expansion in
terms of the coupling $\Gamma$, 
\begin{equation}
\sigma^{\sigma,\sigma}_{\bar{\sigma},\bar{\sigma}}
=
{ \sigma^{\sigma,\sigma}_{\bar{\sigma},\bar{\sigma}} }^{(1)} + {
\sigma^{\sigma,\sigma}_{\bar{\sigma},\bar{\sigma}} }^{(2)} +
\cdots \; .
\end{equation}
%
In lowest order we find the first two diagrams in the right-hand side of 
Fig.~\ref{fig:sfgf}(b) describing spin-dependent quantum charge fluctuations.
From them we obtain the first-order term
\begin{eqnarray}
{
\sigma^{\sigma,\sigma}_{\bar{\sigma},\bar{\sigma}}
}^{(1)}(\varepsilou)
\!\! &=& \!\!
\int \! \rd \varepsilou'
\frac{\gamma^-_\sigma(\varepsilou')}
{\varepsilou+\ri \etb -\varepsilou'+\epsilon_{\bar{\sigma}}}
\nonumber \\
\!\! &+& \!\!
\int \! \rd \varepsilou'
\frac{\gamma^-_{\bar{\sigma}}(\varepsilou')}
{\varepsilou+\ri \etb +\varepsilou'-\epsilon_\sigma}
\nonumber \\
\! &=& \!
\tilde{\gamma}^-_\sigma(\varepsilou+\epsilon_{\bar{\sigma}})
-
\tilde{\gamma}^-_{\bar{\sigma}}(\epsilon_\sigma-\varepsilou)^*,
\label{eqn:sp_selfenergy}
\end{eqnarray}
where
$\gamma^\pm_\sigma \! = \! \sum_r \gamma^\pm_{r \sigma}$ and
$\tilde{\gamma}^\pm_\sigma \! = \! \sum_r \tilde{\gamma}^\pm_{r \sigma}$.
Here we introduced $\tilde{\gamma}^\pm_{r \sigma}$, which is
the Hilbert transform of $\gamma^\pm_{r \sigma}$,
\begin{eqnarray}
& &
\!\!\!\!\!\!\!\!\!
\tilde{\gamma}^\pm_{r \sigma}(\varepsilou)
=
\int \! \rd \varepsilou' \frac{ \gamma^\pm_{r \sigma}
(\varepsilou') } {\varepsilou+\ri \etb-\varepsilou'} \nonumber \\
&=& \frac{\Gamma_{r \sigma}}{2 \pi} \, \rho(\varepsilou-\mu_r)
\biggl \{ \pm \psi \! \left( \frac{1}{2}+\frac{D}{2 \pi T} \right)
\nonumber \\ &\mp& \! {\rm Re} \, \psi \! \left( \frac{1}{2}+ \ri
\frac{\varepsilou-\mu_r}{2 \pi T} \right) +\pi
\frac{\varepsilou-\mu_r}{2 D} \biggl \} - \ri \pi \gamma^\pm_{r
\sigma}(\varepsilou) \; , \;\;\;\;\;
\end{eqnarray}
and $\psi(x)$ denotes the digamma function.
One can expect that the lowest order is sufficient for the 
description of the spin splitting. 
Indeed, for $E_{\rm Z} \!=\!
T \!=\! eV \!=\! 0$ and ferromagnetic leads 
($P_L\!=\!P_R\!=\!P\!>\!0$), 
the spin splitting $\Delta \epsilon$ obtained by 
the scaling theory [Eq.~(\ref{eqn:spliting})] is reproduced, 
$
{
\sigma^{\uparrow,\uparrow}_{\downarrow,\downarrow}
}^{(1)}
(\varepsilou)
\! \approx \!
 \Delta \epsilon$
for $|\varepsilou/\epsilon_0| \! \ll \! 1$. 
We further note that the self-energy given by
Eq.~(\ref{eqn:sp_selfenergy}) also describes the reduction of the
Kondo resonance splitting below the value of the Zeeman
energy $E_{\rm Z}$ for normal (nonmagnetic) leads. For
$P_L\!=\!P_R\!=\!0$ and $T \!=\! eV \!=\! 0$, 
the self-energy then reads,
\begin{equation}
{
\sigma^{\uparrow,\uparrow}_{\downarrow,\downarrow}
}^{(1)}
(\varepsilou)
\approx
-\frac{\Gamma}{\pi}
\tanh^{-1}
\left(
\frac{E_{\rm Z}}{-2 \epsilon_0}
\right),
\;
\Gamma\!=\!\sum_{\sigma,r} \frac{\Gamma_{r \sigma}}{2},
\end{equation}
for 
$\epsilon_\sigma \! < \! 0$ and
$|\varepsilou/\epsilon_{\sigma}| \ll 1$, 
which renormalizes and reduces $E_Z$.
Such a reduction was derived before for the Anderson~\cite{Moore}
and Kondo~\cite{Rosch} models.

\begin{figure}[t]
\includegraphics[width=1.0\columnwidth]{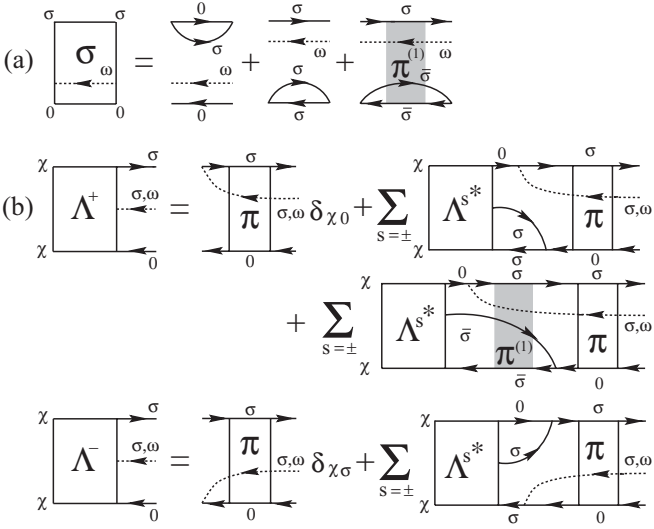}
\caption{ (a) The irreducible self-energy of the reduced
propagator $\pi^{\sigma,\sigma}_{0,0}$. (b) Diagrammatic equations
for the vertex functions ${\Lambda^+}^{\chi,\sigma}_{\chi,0}$ and
${\Lambda^-}^{\chi,\sigma}_{\chi,0}$. An open dotted line connects
to the upper/lower horizontal line for $\Lambda^{\pm}$. Shaded
boxes represent the off-diagonal component of the reduced
propagator in the spin space
${\pi^{\sigma,\sigma}_{\bar{\sigma},\bar{\sigma}}}^{(1)}$. }
\label{fig:vertex}
\end{figure}

Next, we discuss the nature and details of the extended RTA. 
The extension amounts to inserting into all possible
positions in the diagrams of the original RTA expansion 
the lowest-order irreducible self-energy 
${ \sigma^{\sigma,\sigma}_{\bar{\sigma},\bar{\sigma}}
}^{(1)}$ given by Eq.~(\ref{eqn:sp_selfenergy}).
The reduced propagator describing the time evolution of
$P^\sigma_0$ is given by Eq.~(\ref{eqn:rp}) with the replacement of
$\bar{\sigma}$ by $0$:
\begin{equation}
\pi^{\sigma,\sigma}_{0,0}(\varepsilou)
=
1/(\varepsilou+\ri \etb
-\epsilon_\sigma-\sigma^{\sigma,\sigma}_{0,0}(\varepsilou)) \; .
\end{equation}
%
The resulting self-energy of the reduced propagator is
represented by the diagrams 
from Fig.~\ref{fig:vertex}(a). It yields
\begin{eqnarray}
\sigma^{\sigma,\sigma}_{0,0}(\varepsilou) = 
\sum_{s=\pm} \int \! \rd \varepsilou'
\frac{\gamma^s_\sigma(\varepsilou')} {\varepsilou+\ri \etb
-\varepsilou'} \nonumber \\ +  \int \!\! \rd
\varepsilou' \gamma^+_{\bar{\sigma}}(\varepsilou') \, {
\pi^{\sigma,\sigma}_{\bar{\sigma},\bar{\sigma}} }^{(1)}
(\varepsilou-\varepsilou') \;. 
\label{eqn:ch_propagator}
\end{eqnarray}
Here
${
\pi^{\sigma,\sigma}_{\bar{\sigma},\bar{\sigma}}
}^{(1)}
$
is defined by Eq.~(\ref{eqn:sppr}) with the self-energy given by
Eq.~(\ref{eqn:sp_selfenergy}).
The result is similar to the self-energy of the Green function obtained
previously using EOM (Eq.~(8) in Ref.~\cite{Martinek1}). 
The first term of Eq.~(\ref{eqn:ch_propagator}) is related with spin $\sigma$ 
[the first and the second diagrams in the right-hand
side of Fig.~\ref{fig:vertex}(a)]. 
The real part of this term causes a slight shift of the QD-level 
and the imaginary part gives a level broadening. 
The second term [the third diagram in the right-hand
side of Fig.~\ref{fig:vertex}(a)] 
is associated with the Fermi distribution function 
of the opposite spin $\bar{\sigma}$ through $\gamma^+_{\bar{\sigma}}$
and is roughly proportional to 
$\int \rd \omega f^{+}(\omega)
/(\omega \!+\! \ri 0^+ \mp E_Z \mp \Delta \epsilon)$
for $\sigma \!= \uparrow \!\! / \!\! \downarrow$ 
at $eV\!=\!0$. 
Consequently, the second term leads to 
the logarithmic behavior of Kondo correlations.
Our result for
$\sigma^{\sigma,\sigma}_{\bar{\sigma},\bar{\sigma}}$ provides a
microscopic justification to lowest-order expansion
in terms of $\Gamma$ for the 
spin splitting 
$\sigma \Delta \tilde{\epsilon}$ postulated in the EOM scheme
(Eq.~(9) of Ref.~\cite{Martinek1}).

The self-energy of the master equation~(\ref{eqn:master}) is given by
\begin{eqnarray}
\Sigma_{\chi,\sigma} \! &=& \! -2 \ri \, {\rm Im} \! \int \rd
\varepsilou \, \sum_{s=\pm}
{\Lambda^s}^{\chi,\sigma}_{\chi,0}(\varepsilou)
\gamma^s_\sigma(\varepsilou) \; ,
 \label{eqn:selfenergy}
\end{eqnarray}
and $\Sigma_{\chi,0} \! = \! - \! \sum_{\sigma=\uparrow,\downarrow}
\Sigma_{\chi,\sigma}$ ($\chi \! = \! 0, \uparrow, \downarrow$).
The vertex functions $\Lambda^+$ and $\Lambda^-$ are determined by
solving the diagrammatic equations depicted in
Fig.~\ref{fig:vertex}(b). 
In the diagram for $\Lambda^{+/-}$, an open line carrying energy
$\varepsilou$ (dotted line) connects to the upper/lower
horizontal line~\cite{note2}.
The diagrammatic equations can generate diagrams for the resonant
tunneling processes in a recursive manner. The explicit forms are
\begin{eqnarray}
{\Lambda^+}^{\chi,\sigma}_{\chi,0} (\varepsilou)
\!\! &=& \!\!
\pi^{\sigma,\sigma}_{0,0}(\varepsilou) \,
\biggl \{
\delta_{\chi,0}
-
\int \rd \varepsilou'
\sum_{s=\pm}
\frac{
{\Lambda^s}^{\chi,\sigma}_{\chi,0}(\varepsilou')^*
\gamma^s_\sigma(\varepsilou')
}{\varepsilou+ \ri \etb -\varepsilou'}
\nonumber \\
&-& \!\!
\int \rd \varepsilou'
\sum_{s=\pm}
{\Lambda^s}^{\chi,\bar{\sigma}}_{\chi,0}(\varepsilou')^*
\gamma^s_{\bar{\sigma}}(\varepsilou')
{
\pi^{\sigma,\sigma}_{\bar{\sigma},\bar{\sigma}}
}^{(1)}
(\varepsilou-\varepsilou')
\biggl \},
\nonumber \\
\label{eqn:vertex+}
\\
{\Lambda^-}^{\chi,\sigma}_{\chi,0} (\varepsilou)
\!\! &=& \!\!
\pi^{\sigma,\sigma}_{0,0}(\varepsilou)
\nonumber \\
&\times& \!\!
\biggl \{
-
\delta_{\chi,\sigma}
-
\int \! \rd \varepsilou'
\sum_{s=\pm}
\frac{
{\Lambda^s}^{\chi,\sigma}_{\chi,0}(\varepsilou')^*
\gamma^s_\sigma(\varepsilou')
}
{\varepsilou+\ri \etb-\varepsilou'}
\biggl \}.
\label{eqn:vertex-}
\end{eqnarray}
%
Again in the right-hand side of Eq. (\ref{eqn:vertex+}),
we inserted the self-energy
${\sigma^{\sigma,\sigma}_{\bar{\sigma},\bar{\sigma}}}^{(1)}$
via
${\pi^{\sigma,\sigma}_{\bar{\sigma},\bar{\sigma}}}^{(1)}$.

From $P^{\rm st}_{\sigma}$, which is obtained 
by solving the master equation Eq.~(\ref{eqn:master}),
we find the average occupation $N$ and the local magnetization $M$ of the
QD, 
\begin{eqnarray}
N=P^{\rm st}_\uparrow+P^{\rm st}_\downarrow \; ,
\; \;
M=(1/2)
(P^{\rm st}_\uparrow-P^{\rm st}_\downarrow) \; .
\end{eqnarray}
The spin-dependent nonequilibrium DOS of the QD is defined as
\begin{eqnarray}
\rho_\sigma(\varepsilou)
\! = \!
-
\frac{1}{2 \ri \pi}
\{
G^>_\sigma(\varepsilou)-G^<_\sigma(\varepsilou)
\},
\label{eqn:spinDOS}
\end{eqnarray}
where the lesser/greater Green functions are expressed, using
Eqs. (\ref{eqn:vertex+}) and (\ref{eqn:vertex-}), as
\begin{equation}
G^{\gtrless}_\sigma (\varepsilou)
=
2 \ri \, {\rm Im} \!\!\!
\sum_{\chi=0,\uparrow,\downarrow}
P^{\rm st}_\chi
{\Lambda^{\pm}}^{\chi,\sigma}_{\chi,0} (\varepsilou).
\label{eqn:green_functions}
\end{equation}
%
Also the transport current flowing into the QD through the junction
$r$, $I_r$, can be expressed in terms of the Green
functions. The spin-resolved components read
\begin{equation}
I_{r \sigma}
=
-
\ri e
\int \! \rd \varepsilou \,
\{
\gamma_{r \sigma}^+ (\varepsilou) \,
G^{>}_\sigma (\varepsilou)
+
\gamma_{r \sigma}^- (\varepsilou) \,
G^{<}_\sigma (\varepsilou)
\}.
\label{eqn:current}
\end{equation}

One can prove that the approximation satisfies
spin and charge conservation~\cite{note4}.
Equations
(\ref{eqn:selfenergy}),
(\ref{eqn:green_functions})
and
(\ref{eqn:current})
combine to
$
\sum_r I_{r \sigma}
\! = \!
\ri e
\sum_\chi
P^{st}_\chi \,
\Sigma_{\chi, \sigma}
$,
which 
by virtue of the master equation, Eq.~(\ref{eqn:master}) 
leads to the spin-resolved current conservation, 
$\sum_r I_{r \sigma} \! = \! 0$.
The conservation laws for charge and spin,
\begin{eqnarray}
I_{L} \! =\! -I_{R} 
\!=\! I,
\; \; \;
\sum_r
(I_{r \uparrow} - I_{r \downarrow})
\! = \! 0,
\label{eqn:conservation}
\end{eqnarray}
follows the spin-resolved current conservation. 
Equation (\ref{eqn:conservation}) is valid only for 
collinear magnetizations, where the spin is
conserved at each vertex in the diagrams of the self-energy for the
master equation, regardless of the sign of $P_{L(R)}$. 
This fact ensures that the self-energy of master equation 
does not mix diagonal and off-diagonal states 
and 
allows one to formulate a closed equation for the 
diagonal components of the reduced density matrix.

The linear conductance, 
$G_{\sigma}^{\rm lin} \! \equiv \! \lim_{V
\rightarrow 0} d I_\sigma(V)/d V$ 
[$I_{\sigma} \!=\! I_{r \sigma}$], probes the equilibrium spin-dependent
DOS, $\rho_\sigma^{eq}(\omega)$, around the Fermi energy:
\begin{eqnarray}
\frac{G_\sigma^{\rm lin}}{G_{\rm K}}
=
2 \pi
\frac{\Gamma_{L \sigma} \Gamma_{R \sigma}}
{\Gamma_{L \sigma} \! + \! \Gamma_{R \sigma}}
\int \! \rd \varepsilou \,
\biggl \{- \frac{\partial f^+(\varepsilou)}{\partial \varepsilou} \biggl \}
\,
\rho_\sigma^{eq} (\varepsilou),
\label{eqn:conductance}
\end{eqnarray}
where $G_{\rm K} \equiv e^2/(2 \pi \hbar)$ is the conductance quantum. We
symmetrized the expression
Eq.~(\ref{eqn:conductance}) using the spin and charge
conservation, Eq.~(\ref{eqn:conservation}).

\section{Results and discussions}
\label{sec:result}

In this section, we will present results
obtained by solving the integral equations numerically. 
In all results
we tested the numerical accuracy by checking
the spin and charge conservation
(
the relative error is smaller than
$5 \! \times \! 10^{-6}$).
We also checked the spectral sum rule, 
$
\int \rd \varepsilou
\,
\rho_\sigma (\varepsilou)
\!=\!
P^{\rm st}_0+P^{\rm st}_\sigma
$,
which follows from the fact that
${\Lambda^{\pm}}^{\chi,\sigma}_{\chi,0} (\varepsilou)$
is analytic in the upper half plane
and decays at infinity as $1/\varepsilou$. 
The sum rule is always satisfied with accuracy
better than $5 \! \times \! 10^{-4}$. 

In Sec.~\ref{sec:splitting}, we will discuss the splitting of the
zero-bias anomaly.
Subsections~\ref{sec:restoration} and \ref{sec:nonlinear} 
provide a discussion of the restoration of the Kondo
resonance as well as of properties of the nonequilibrium Kondo
effect.
In theses subsections, we will assume a left-right symmetry in the
coupling strengths 
$\Gamma_L \! =\! \Gamma_R$ 
[$\Gamma_r \! \equiv \! (\Gamma_{r \uparrow}+\Gamma_{r \downarrow})/2$] 
and 
spin polarization factors $P_L \! =\! |P_R|$, and  $a_c \! = \! 0$.
In equilibrium these symmetries are not important because, as one
can see from a unitary transformation~\cite{Martinek2}, 
an asymmetry gives just a constant prefactor to the linear
conductance. However, this is not the case for a nonequilibrium state,
as will be shown in Sec.~\ref{sec:asym}.

\subsection{Splitting of the zero bias anomaly}
\label{sec:splitting}

\begin{figure}[t]
\includegraphics[width=0.8\columnwidth]{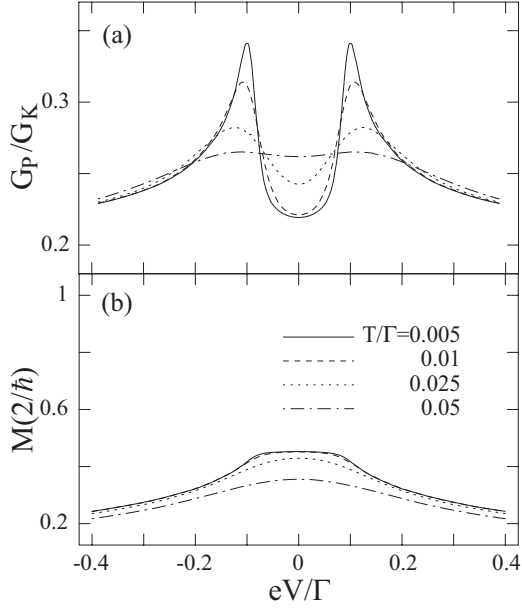}
\caption{ (a) The nonlinear differential conductance and (b) the
local magnetization $M$ for the P alignment ($P_L \! = P_R \! = \!
0.1$) and for symmetric coupling $\Gamma_L \! = \! \Gamma_R$.
Solid, dashed, dotted and dot-dashed lines are results for various
temperatures $T/\Gamma=0.005,0.01,0.025$, and $0.05$,
respectively.
Parameters are:
$\epsilon_0/\Gamma \!=\! -2$,
$D/\Gamma \! =\! 50$.
}
\label{fig:temperature}
\end{figure}

Figure~\ref{fig:temperature}(a) shows the nonlinear differential
conductance $G_{\rm P}(V) \! \equiv \! dI(V)/dV$ 
for P alignment and symmetric
coupling, $\Gamma_L \! =\! \Gamma_R$, and spin polarization,
$P_L \! =\! P_R$. The splitting of the zero-bias anomaly is
well observable for low temperatures.
The enhancement of the
peak at low temperatures is a clear sign of Kondo correlations.
The value of the splitting is independent of temperature and
almost coincides with the prediction of the scaling theory
Eq.~(\ref{eqn:spliting}), $2 \Delta \epsilon \! \approx \! 0.204 \Gamma$.
At high temperature, the two peaks are smeared out. 
However, the magnitude of the splitting 
is not affected by the temperature. 

Figure~\ref{fig:temperature}(b) shows the corresponding plot for
the local magnetization $M$. For 
small bias voltage $|eV|
\lesssim |\Delta \epsilon|$, it increases with decreasing temperature.
At the same time, the differential conductance decreases
[Fig.~\ref{fig:temperature}(a)]. This indicates that the suppression
of the zero-bias anomaly can be attributed to the quenching of the spin-flip
scattering at low temperatures.
When the applied bias voltage exceeds
the spin-splitting energy,
$|eV| \gtrsim |\Delta \epsilon|$,
a spin-flip channel opens.
As a result, the local magnetization starts to decrease
and peaks appear in the nonlinear differential conductance.

\subsection{Restoration of Kondo resonance}
\label{sec:restoration}

A remarkable feature of the spin splitting induced by 
ferromagnetism is that it can be compensated by an applied magnetic
field~\cite{Martinek1,Martinek2}. At the point of full
compensation, the strong coupling limit can be retrieved. 
The Kondo temperature in the leading logarithmic approximation is
given by~\cite{Martinek1}
\begin{eqnarray}
T_{\rm K}(P) \approx \tilde{\epsilon} \, \exp \biggl \{ \frac{-\pi
\tilde{\epsilon}}{\Gamma} \, \frac{ \arctan(P)}{P} \biggl \} \; .
\label{eqn:Kondo}
\end{eqnarray}
Here we considered the case $P_L\!=\!P_R\!=\!P$.
The ratio $\Gamma/(\pi \tilde{\epsilon})$ corresponds the value
$J_0(\nu_\uparrow+\nu_\downarrow)$ of Ref.~\cite{Martinek1}. The
cutoff energy $\tilde{\epsilon}$ follows from
$\tilde{\epsilon} \! = \! -\epsilon_0 \! + \! \Gamma/(2 \pi) \ln
(\tilde{\epsilon}/D)$.

\begin{figure}[t]
\includegraphics[width=0.75\columnwidth]{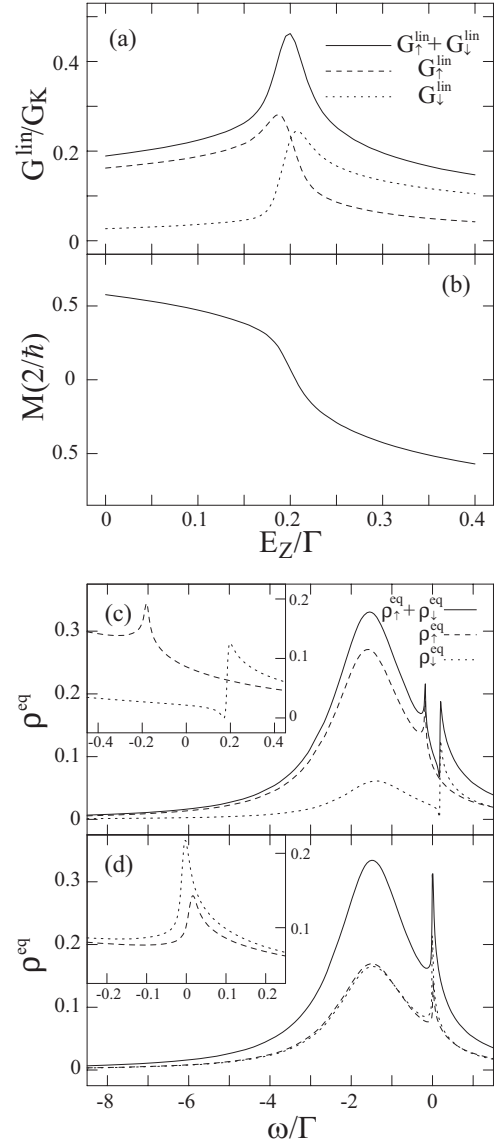}
\caption{ (a) The linear conductance $G^{\rm lin}$ and the local
magnetization $M$ as a function of the Zeeman energy $E_{\rm Z}$
for the P alignment ($P_L \!=\! P_R \!=\! 0.2$). In panel (a), the
spin-resolved conductance is plotted by dashed (dotted) line
for up (down) spin.
The equilibrium DOS for (c) $E_{\rm Z}/\Gamma \! = \! 0$ and (d)
$E_{\rm Z}/\Gamma \! = 0.205$. 
Solid and dashed (dotted) lines show the total DOS 
and the spin resolved DOS for up (down) spin. Insets in these
panels are blow-ups of the Kondo resonance.
Parameters are:
$T/\Gamma \!=\! 0.005$,
$\epsilon_0/\Gamma \!=\! -2$,
$D/\Gamma \! =\! 50$.
}
\label{fig:linear}
\end{figure}

Figure~\ref{fig:linear}(a) and (b) show the linear conductance
$G^{\rm lin}$ and the local magnetization $M$ as a function of
Zeeman energy $E_{\rm Z}$ in equilibrium for the parallel alignment. The
average occupation $N$ depends slightly on $E_{\rm Z}$ and varies
between $N \! \approx \! 0.88$ and $N \! \approx \! 0.91$.
At $E_{\rm Z}\!=\!0$, the direction of the local magnetization $M$ and
lead magnetizations are aligned. With increasing Zeeman
energy the local magnetization $M$ changes the direction.
At a peak in the linear conductance $G^{\rm lin}$, where spin-flip
processes are expected to be enhanced [solid line in
Fig.~\ref{fig:linear}(a)], we observe the local magnetization
$M\!=\!0$ in agreement with results from Ref.~\cite{Martinek2}.
The spin state of the QD is responsible for the slight asymmetry
in the conductance around the $M\!=\!0$ point. In the left (right)
side of the peak, because the QD level is occupied by an 
up-spin (down-spin) electron, the co-tunneling current for up-spin (down-spin)
component is dominant [dashed (dotted) line in
Fig.~\ref{fig:linear}(a)]. As the up-spins are majority spins, a
larger conductance is obtained on the left side of the peak.


We analyzed the DOS at 
$E_{\rm Z}\!=\!0$ 
[Fig.~\ref{fig:linear}(c)] 
and at 
$E_{\rm Z}\!=\!0.205 \, \Gamma$, 
where $M\! \approx \!0$ [Fig.~\ref{fig:linear}(d)]. 
The solid and dashed (dotted) lines 
show the total DOS $\rho(\varepsilou) \equiv \sum_\sigma
\rho_\sigma(\varepsilou)$ and the up-spin (down-spin) component of the
DOS $\rho_\sigma(\varepsilou)$, respectively. 
Figure~\ref{fig:linear}(c) and its inset demonstrate that the 
Kondo resonance splits at $E_{\rm Z}\!=\!0$. 
The two resonances, which appear rather well resolved, should be
smeared by decoherence effects. These effects are beyond the present
approximation, as we will discuss in Sec.~\ref{sec:decoherence}.
The enhancement/suppression of the 
up/down spin (dashed/dotted line) component of 
the QD-level DOS 
(broad peak around $\varepsilou/\Gamma\!=\!-2$) 
indicates the positive local magnetization,
$M \!>\!0$.

At $E_{\rm Z}\!=\!0.205 \, \Gamma$, the Kondo resonance is restored at
the Fermi energy [solid line in Fig.~\ref{fig:linear}(d)]. 
We can observe that the resonance for the minority-spin component
(dotted line) is higher than that of the the majority-spin
component (dashed line) [inset in Fig.~\ref{fig:linear}(d)]. 
Loosely speaking, it is because the
minority (majority) spin in the QD is well (poorly) screened by
majority (minority) spin in leads. 
This tendency is consistent 
with the zero-temperature result of NRG~\cite{Martinek2} and 
the Friedel sum rule, which predicts $\rho_{\sigma}(0) \sim
1/\Gamma_{\sigma}$~\cite{Martinek2}.

As for the QD-level DOS,
up-spin and down-spin components are almost the same
(at $E_{\rm Z}\!=\!0.205 \, \Gamma$), 
which indicates $M \! =\!0$
[dotted and dashed lines in Figs.~\ref{fig:linear}(d)]. 
However, though the QD level can be spin polarized depending on $E_Z$, 
the total QD-level DOS is not affected by $E_Z$, 
i.e. the solid lines in Figs.~\ref{fig:linear}(c) and (d) 
can almost overlap around the broad peak. 
Such a feature is also obtained from the NRG work~\cite{MSChoi}.

\subsection{Kondo resonance out of equilibrium}
\label{sec:nonlinear}

\begin{figure}[t]
\includegraphics[width=1.0\columnwidth]{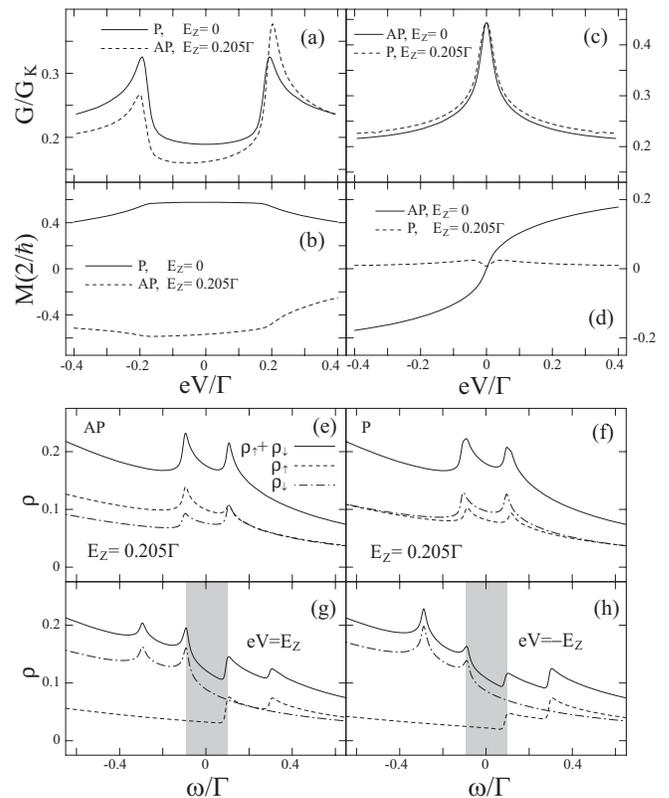}
\caption{ The nonlinear differential conductance $G$, (a) and (c),
and the local magnetization $M$, (b) and (d), for a symmetric
coupling $\Gamma_L\!=\!\Gamma_R$.
Solid lines and dashed lines in (a) and (c) are plots for $E_{\rm
Z} \! =\! 0$ and $E_{\rm Z} \! =\!0.205 \, \Gamma$, respectively. The
solid lines in (a) and (c) are for the P alignment ($P_L \! = \!
P_R \! = \! 0.2$) and the AP alignment ($P_L \! = \! -P_R \! = \!
0.2$)], respectively.
The dashed lines in (a) and (c) are for the AP alignment and the P
alignment, respectively.
The average occupation is $N \! \approx \! 0.89$ for all parameter
regimes.
The panels (e) and (f) show the corresponding DOS
$\rho(\varepsilou)$ of the solid and the dashed lines in panel
(c) at $eV/\Gamma \!=\! 0.205$. The dashed and dot-dashed lines
are for $\rho_\uparrow(\varepsilou)$ and
$\rho_\downarrow(\varepsilou)$, respectively.
Panels (g) and (h) show 
plots of the DOS $\rho(\varepsilou)$ of the dashed line in panel
(a) at $eV \! = \! E_{\rm Z}$ (g), and at $eV \! = \! -E_{\rm Z}$
(h).
The other parameters are the same as for
Fig.~\ref{fig:linear}. } 
\label{fig:nonlinear}
\end{figure}

In this subsection, we will discuss the nonequilibrium Kondo effect.
Fig.~\ref{fig:nonlinear} shows the nonlinear differential
conductance, $G$ [(a) and (c)], and the local magnetization, $M$ [(b)
and (d)], for symmetric coupling. As discussed in
Sec.~\ref{sec:splitting}, we can observe the splitting of the
zero-bias anomaly for parallel alignment [solid line in (a)].
However, for antiparallel alignment [solid line in (c)], the splitting
vanishes. Moreover, for positive (negative)
bias voltage spin accumulates in the QD with
positive (negative) local magnetization [solid line in (d)].

As discussed in Sec.~\ref{sec:restoration}, the zero-bias
anomaly is expected to be restored when the spin splitting
is compensated by an applied magnetic field. 
The dashed line in (c)
shows the case for the P alignment with $E_{\rm Z}\!=\!0.205 \, \Gamma$.
Figures~\ref{fig:nonlinear} (e) and (f) show the nonequilibrium DOS
$\rho_\sigma(\varepsilou)$ for the AP alignment and that for the P
alignment with the Zeeman energy $E_{\rm Z}\!=\!0.205 \, \Gamma$. The
total DOS has two peaks (solid lines) 
at two chemical potentials, similar in shape to that of a 
nonequilibrium Kondo DOS split by an applied voltage.
However, now the two peaks are spin polarized, 
i.e. there is clear difference in the weights for the spin-up
 component $\rho_\uparrow(\varepsilou)$ (dashed lines) and the
spin-down component $\rho_\downarrow(\varepsilou)$ (dot-dashed
lines).

For the AP alignment with the Zeeman energy 
$E_{\rm Z}\!=\!0.205 \, \Gamma$, 
the splitting persists [dashed line in (a)], but
in this case the splitting is caused by the Zeeman
energy~\cite{Meir1}, and a strong asymmetry between two peaks is
obtained.
Intuitively, the result can be understood by the following argument:
In the regime $|eV| \! \lesssim \! E_{\rm Z}$,
a spin-down occupies the QD level.
Above a positive bias voltage
$eV \! \ge \! E_{\rm Z}$, 
spin-flip tunneling events from
a left lead spin-up state to a right lead spin-down state contribute to
the current.
In this case, because both initial and final states are majority-spin states,
a large peak is expected.
On the other hand, at negative bias voltages
$eV \! \le \! -E_{\rm Z}$, the spin-flip tunneling processes are
between minority spin states. Hence
the peak in the differential conductance is weak.

A more precise explanation taking into account Kondo correlations can be given
based on the spin-polarized nonequilibrium Kondo resonance.
Figure~\ref{fig:nonlinear}(g) shows $\rho(\varepsilou)$
for $eV \! = \! E_{\rm Z} \! = \! 0.205 \, \Gamma$. The two spin
polarized Kondo resonances out of equilibrium split further into
four peaks (solid line) by the energy $eV$~\cite{Meir1}. The lower
two peaks at $-E_{\rm Z} + eV/2$ and $-E_{\rm Z} - eV/2$ are for
spin-down (dot-dashed line), related to the left and right lead 
chemical potential at $\pm eV/2$, respectively. The peak
at $-E_{\rm Z} + eV/2$ is larger than the peak at $-E_{\rm Z} -
eV/2$, because the spin-down electron in the QD can be screened by majority
spins of the left lead with spin-up.
The peaks at $E_{\rm Z} \pm eV/2$ are for spin-up electrons (dashed line),
and a larger amplitude of the peak at $E_{\rm Z} - eV/2$ can be
explained in the same way as above.
Under the condition $eV \! = \! E_{\rm Z}$, the positions of 
the two large peaks at $E_{\rm Z} - eV/2$ and $- E_{\rm Z} + eV/2$
coincide with the two chemical potentials. Thus, tunneling
electrons with energy between the left and right chemical potentials
[shaded region in (g)] can profit from the larger DOS inside the QD.
Consequently, the differential conductance $G_{\rm AP}(V)$ at
positive bias voltage $eV=E_{\rm Z}$ is enhanced more strongly.
Finally, Fig.~\ref{fig:nonlinear}(h) shows the DOS $\rho(\varepsilou)$
for $eV \! = \! -E_{\rm Z} \! = \! -0.205 \, \Gamma$. By repeating
the same arguments as presented above, we can explain the fact
that at negative bias voltage $eV=-E_{\rm Z}$ the DOS
$\rho(\varepsilou)$ of the QD at energies 
between the left and right Fermi levels
[shaded region in (h)] is lower, and consequently the peak of
the differential conductance $G_{\rm AP}(V)$ is smaller as well.

\begin{figure}[t]
\includegraphics[width=0.8\columnwidth]{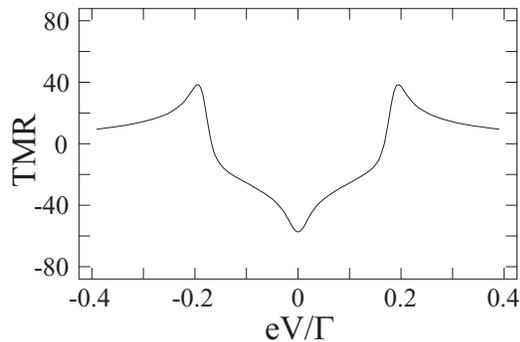}
\caption{ The tunnel magnetoresistance TMR as function
 of the bias voltage $V$. Spin polarization factors are $P_L \! =\!
|P_R| \! =\! 0.2$.
The further parameters are the same as for Fig.~\ref{fig:linear}.
}
\label{fig:TMR}
\end{figure}

The anomalous behavior of the nonlinear conductance determines the
shape of the TMR. Figure~\ref{fig:TMR} presents the TMR defined as
\begin{eqnarray}
{\rm TMR} \equiv (G_{\rm P}(V)-G_{\rm AP}(V))/G_{\rm AP}(V).
\label{eqn:tmr}
\end{eqnarray}
We find a large {\it negative} TMR at zero bias. This is in
contrast to the TMR of ferromagnetic tunnel junctions, where one finds 
usually a maximum of the TMR at zero bias and then a decrease
with increasing bias voltage $V$~\cite{Maekawabook,Utsumi1}.

\subsection{Left-right asymmetry in system parameters}
\label{sec:asym}

In the previous subsection we showed that the spin-polarization of
the Kondo resonance and the spin-accumulation 
for a nonequilibrium situation depend on the
orientation of the lead magnetizations. We can expect that a
left-right asymmetry in coupling strengths, $\Gamma_R \! \neq \!
\Gamma_L$, and  spin polarization, $P_R \! \neq \! P_L$, or an
asymmetry in the applied voltage 
could  affect these properties as well and leave footprints on the
nonlinear differential conductance. In the following we discuss 
several examples.

\begin{figure}[t]
\includegraphics[width=1.0\columnwidth]{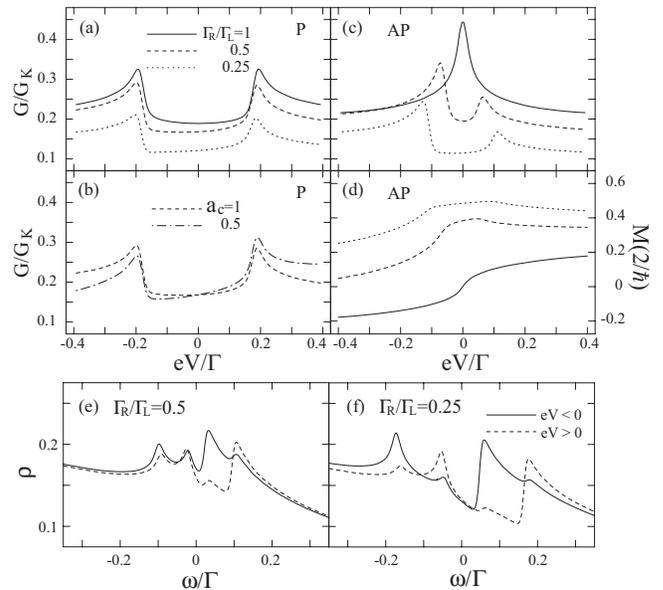}
\caption{ (a) The differential conductance for the
P alignment $G_{\rm P}(V)$ ($P_L \! = P_R \! = \! 0.2$). 
Solid lines, dashed
lines and dotted lines are results for $\Gamma_R/\Gamma_L \! = \!
1, 0.5$, and $0.25$. (b) The differential conductance $G_{\rm
P}(V)$ for $\Gamma_R/\Gamma_L \! =\! 0.5$ with the asymmetry factor
$a_c \! = \! 0$ (dashed line) and $0.5$ (dot-dashed line).
(c) The differential conductance with the AP alignment 
$G_{\rm AP}(V)$ ($P_L\! = -P_R \! = \! 0.2$) 
and (d) the nonequilibrium local magnetization $M$.
(e), (f) The total DOS $\rho(\varepsilou)$ in the AP alignment: 
(e) The solid (dashed) line is for $\Gamma_R/\Gamma_L \! = \! 0.5$ 
at $eV/\Gamma \!=\! -0.08$ ($0.07$). 
(f) The solid (dashed) line is for $\Gamma_R/\Gamma_L \! = \! 0.25$ 
at $eV/\Gamma \!=\! -0.13$ ($0.11$).
The other parameters are the same as for
Fig.~\ref{fig:linear}. } 
\label{fig:asymmcoupling}
\end{figure}

Figure~\ref{fig:asymmcoupling}(a) shows the nonlinear differential
conductance $G_{\rm P}(V)$ for the P alignment ($P_L \! = \! P_R
\! = \! 0.2$) for varying values of the ratio of coupling strengths
$\Gamma_R/\Gamma_L$, while keeping the total coupling 
$\Gamma = \Gamma_L + \Gamma_R$ unchanged. 
The values of the splittings are almost the same for all cases. 
The reason is the following: 
The spin splitting Eq.~(\ref{eqn:spliting}) is expressed as
the sum of the left and right contributions
$\Delta \epsilon \equiv \Delta \epsilon_L+\Delta \epsilon_R$ 
(one can obtain the relation for $\Delta \epsilon_r $ using
Eq.~(\ref{eqn:spliting}) and adding a lead index $r$ to the
$\Gamma$ and $P$). 
For $P_L \! = \! P_R \! = \! P$, the value of the splitting $\Delta \epsilon$ 
is proportional to $P (\Gamma_L + \Gamma_R)$, 
which is independent of the ratio of coupling strength. 
The asymmetric capacitive coupling does not alter the conductance
$G(V)$ by much. Figure~\ref{fig:asymmcoupling}(b) shows curves for
the asymmetry factor $a_c \! = \! 0$ (dashed line) and $0.5$
(dot-dashed line) for $\Gamma_R/\Gamma_L \! =\! 0.5$ [the dashed
line is the same as that in (a)]. The asymmetry factor just tilts
the curve slightly.

Figure~\ref{fig:asymmcoupling}(c) presents the nonlinear
conductance $G_{\rm AP}(V)$ for the AP alignment for
different values of
$\Gamma_R/\Gamma_L$, and $a_c \! = \! 0$. 
For an asymmetric coupling, we find again a 
splitting of the zero bias anomaly.
Such a splitting was observed experimentally~\cite{Pasupathy}
(details are given in Sec.~\ref{sec:exp}). 
It can be understood because
the spin splitting due to the exchange interaction with the left
$\Delta \epsilon_L$ and right lead $\Delta \epsilon_R$ cannot
compensate each other $\Delta \epsilon_L \neq - \Delta \epsilon_R $
for $\Gamma_L \! \neq \! \Gamma_R$. Actually, with decreasing
$\Gamma_R/\Gamma_L$, $\Delta \epsilon$ increases and thus the
splitting is enhanced.

We observe the asymmetry in the two peaks' amplitude similar to the
case of the AP alignment with the Zeeman splitting [dashed line in
Fig.~\ref{fig:nonlinear}(a)]. Actually, this asymmetry can be
explained almost in the same way. The only difference is that the
large peak is at negative bias voltage [dashed and dotted lines in
Fig.~\ref{fig:asymmcoupling}(c)] as opposed to the dashed
line in Fig.~\ref{fig:nonlinear}(a), because, $\Delta
\epsilon$ is negative whereas the Zeeman energy is positive.
The solid (dashed) lines in panels (e) and (f) show the total DOS
$\rho(\omega)$ for the AP alignment at negative (positive) bias
voltage. We can check that the larger DOS at negative bias voltage
(solid lines) is responsible for higher conductance peak.

\begin{figure}[t]
\includegraphics[width=1.0\columnwidth]{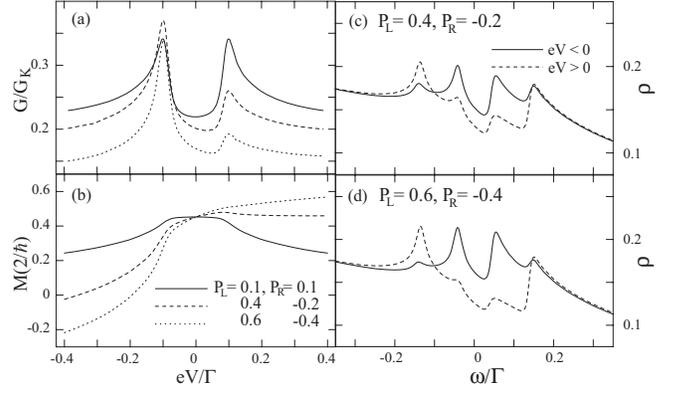}
\caption{ (a) The nonlinear conductance $G(V)$ and (b) the local
magnetization $M$ for various spin polarizations ($P_L$,
$P_R$) and for symmetric coupling $\Gamma_R \!=\! \Gamma_L$.
The average value of the polarization factors in the left and right leads
are fixed as $(P_L+P_R)/2 \!=\! 0.1$. The solid, dashed, and dotted
lines are for $P_L \! = \! 0.1$, $0.4$, and $0.6$, respectively.
The total DOS $\rho(\varepsilou)$ at $eV/\Gamma \!=\! -0.1$ (solid
line) and $eV/\Gamma \!=\! 0.1$ (dashed line) for $P_L \!=\! 0.4$
(c) and $P_L \!=\! 0.6$ (d).
The other parameters are the same as for Fig.~\ref{fig:linear}.
}
\label{fig:asymP}
\end{figure}

Next, we will look at consequences of an asymmetry in left
and right spin polarization factors $P_L \! \neq \! P_R$. The
dashed and dotted lines in Fig.~\ref{fig:asymP} show the
differential conductance $G(V)$ (a) and local magnetization $M$
(b) for various spin polarization factors but for symmetric
coupling strength $\Gamma_R = \Gamma_L$. For each curve, the
average value of polarization factors in the left and right leads was
kept constant $(P_L+P_R)/2 \!=\! 0.1$. The solid lines show the
result for the equal polarizations ($P_L \! = \! P_R \!=\!
0.1$). As the spin polarization of the left lead increases ($P_L \!
= \! 0.4$ for dashed lines, $P_L \! = \! 0.6$ for dotted lines)
the local magnetization increases (decreases) for positive
(negative) bias voltage [Fig.~\ref{fig:asymP}(b)] and accordingly,
the difference between the two peaks grows [Fig.~\ref{fig:asymP}(a)].
The solid (dashed) lines in panels (c) and (d) are the total DOS 
at negative (positive) bias voltage [(c) for
$P_L\!=\!0.4$ and (d) for $0.6$]. It confirms that with increasing
$P_L$, the difference in the intensity increases, causing the large
difference in peak heights.

\section{Decoherence effect}
\label{sec:decoherence}

In this section, we will turn our attention to the effects of 
decoherence. Since the
transverse spin relaxation time, $T_2$, is defined as the decay time
of the off-diagonal components of the reduced density matrix
$P^\sigma_{\bar{\sigma}}$, one might identify it within 
the Markov approximation with 
the imaginary part of 
the corresponding off-diagonal component of the self-energy
in Fourier space,
\begin{equation}
1/T_2
=
-{\rm Im} \,
\sigma^{\sigma,\sigma}_{\bar{\sigma},\bar{\sigma}}
(\epsilon_\sigma \! - \! \epsilon_{\bar{\sigma}}).
\end{equation}
The relevant frequency in the argument 
$\epsilon_\sigma \! - \! \epsilon_{\bar{\sigma}}$ is the
energy difference between the states with spin $\sigma$ 
and the opposite spin $\bar{\sigma}$. 
From Eq. (\ref{eqn:sppr}), one can
conclude, that $1/T_2$ is closely related to the broadening of Kondo
resonance. When evaluating the self-energy we note that the  
lowest-order contribution to
$\sigma^{\sigma,\sigma}_{\bar{\sigma},\bar{\sigma}}$ does not contain an
 imaginary part. In the following, we therefore, perform the second order
perturbation expansion in terms of $\Gamma$ at zero
temperature $T\!=\!0$ and $\epsilon_{r \sigma} \! \equiv \!
\epsilon_{\sigma}-\mu_{r} <0$.

\begin{figure}[t]
\includegraphics[width=0.8\columnwidth]{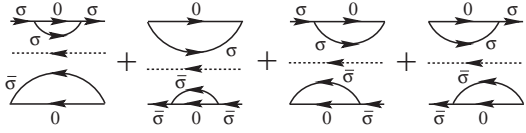}
\caption{ 
The four additional diagrams of second order expansion
for the irreducible self-energy
$\sigma^{\sigma,\sigma}_{\bar{\sigma},\bar{\sigma}}$. 
One bubble dresses each horizontal line. The other two diagrams, where one
bubble is inside the other bubble are depicted in
Fig.~\ref{fig:sfgf}(b). 
Here, dashed lines carry the energy
$\epsilon_\sigma-\epsilon_{\bar{\sigma}}$. 
}
\label{fig:2ndorder}
\end{figure}

The first diagram of the second-order expansion, 
depicted in Fig.~\ref{fig:sfgf}(b), is evaluated as
\begin{eqnarray}
{\sigma^{\sigma,\sigma}_{\bar{\sigma},\bar{\sigma}}}^{\text{(2-1)}}
(\epsilon_\sigma \! - \! \epsilon_{\bar{\sigma}})
= 
\sum_{\sigma'}
\int \!
\rd \varepsilou'
\rd \varepsilou''
\frac{1}
{\epsilon_{\sigma}-\omega'+\ri \etb}
\nonumber \\
\times
\frac{
\gamma^+_{\sigma'}(\varepsilou'')
\,
\gamma^-_{\sigma}(\varepsilou')
}{
(
\epsilon_{\sigma}
-\epsilon_{\sigma'}
-\varepsilou'+\varepsilou''
+\ri \etb)
(\epsilon_{\sigma}-\varepsilou'+\ri \etb)}. 
\label{eqn:sp_selfenergy_2nd}
\end{eqnarray}
For simplicity, we assume
here a flat DOS with cutoff energy $D$, i.e., we
replace the Lorentzian cutoff function by $\rho_{\rm
c}(\varepsilou) \rightarrow \theta(\varepsilou+D)
\theta(D-\varepsilou)$ with $|\epsilon_{r \sigma}| \! \ll \! D$. 
Then, the imaginary part of Eq.
(\ref{eqn:sp_selfenergy_2nd}) reads,
\begin{eqnarray}
&-&\!{\rm Im} \,
{\sigma^{\sigma,\sigma}_{\bar{\sigma},\bar{\sigma}}}^{\text{(2-1)}}
(\epsilon_\sigma \! - \! \epsilon_{\bar{\sigma}}) 
\nonumber \\
&\approx& \!\!\!\!\! 
\sum_{\sigma' r r'} 
\pi 
\frac{\Gamma_{r \sigma} \Gamma_{r' \sigma'}}{(2 \pi)^2} 
\frac{ \epsilon_{r \sigma}-\epsilon_{r' \sigma'} }
{\epsilon_{r \sigma} \, \epsilon_{r' \sigma'}} 
\, 
\theta (\epsilon_{r \sigma}-\epsilon_{r' \sigma'}) \;.
\end{eqnarray}
The second diagram of the second-order expansion
is obtained from Eq.~(\ref{eqn:sp_selfenergy_2nd}) as
${\sigma^{\sigma,\sigma}_{\bar{\sigma},\bar{\sigma}}}^{\text{(2-2)}}
(\varepsilou)
\! = \! -
{\sigma^{\bar{\sigma},\bar{\sigma}}_{\sigma,\sigma}}^{\text{(2-1)}}
(-\varepsilou)^*$.
%
In addition to the above two diagrams there are four other diagrams
(Fig.~\ref{fig:2ndorder}); however their
imaginary part vanishes for $\epsilon_{r
\sigma} \! < \! 0$:
\begin{eqnarray}
& &
 {\rm Im}\,
{\sigma^{\sigma,\sigma}_{\bar{\sigma},\bar{\sigma}}}
^{\text{(2-3)}}
(\epsilon_\sigma \! - \! \epsilon_{\bar{\sigma}})
=
\nonumber \\ & & {\rm Im} \int \! \rd \varepsilou' \rd
\varepsilou'' \frac{ \gamma^-_{\bar{\sigma}}(\varepsilou') \,
\gamma^-_{\sigma}(\varepsilou'') } {\varepsilou'-\varepsilou''
+\epsilon_{\sigma}-\epsilon_{\bar{\sigma}} +\ri \etb} 
\nonumber \\
&\times& 
\biggl (
\frac{1}{\varepsilou'-\epsilon_{\bar{\sigma}}+\ri \etb} 
+
\frac{1}{\epsilon_{\sigma}-\varepsilou''+\ri \etb} \biggl
)^2 
\nonumber \\ &\approx& 
\sum_{r r'} \frac{ \Gamma_{r \sigma} \,
\Gamma_{r' \bar{\sigma}} }{4 \pi} \biggl (
\frac{\theta(\epsilon_{r' \bar{\sigma}})}{\epsilon_{r \sigma}} +
\frac{\theta(\epsilon_{r \sigma})} {\epsilon_{r' \bar{\sigma}}}
\biggl ) =0 \; .
 \label{eqn:sp_selfenergy_bubble}
\end{eqnarray}
%
By adding the contributions from
${\sigma^{\sigma,\sigma}_{\bar{\sigma},\bar{\sigma}}}^{\text{(2-1)}}$
and
${\sigma^{\sigma,\sigma}_{\bar{\sigma},\bar{\sigma}}}^{\text{(2-2)}}$,
one obtains the transverse spin relaxation rate in 2nd order,
\begin{eqnarray}
\frac{1}{T_2^{(2)}} \approx \sum_{\sigma \sigma'}
\frac{\Gamma_{L \sigma} \Gamma_{R \sigma'} \, |\epsilon_{L
\sigma}-\epsilon_{R \sigma'}|} {4 \pi \, \epsilon_{L \sigma} \,
\epsilon_{R \sigma'}} \; .
\label{eqn:t2}
\end{eqnarray}
The rate is proportional to the sum of spin flip and spin preserving 
co-tunneling currents.
This result is consistent with $\sum_\sigma 1/\tau_{\sigma}$ 
derived in Refs.~\cite{Meir1,Meir2} 
($\tau_{\sigma}$ is defined in Eq. (5) of
Ref.~\cite{Meir1} or Eq. (43) of Ref.~\cite{Meir2}), where 
the decoherence effect due to the dissipative current
flow for finite bias and/or magnetic field even at zero temperature
was first pointed out.

Because the RTA ignores
$\sigma^{\sigma,\sigma}_{\bar{\sigma},\bar{\sigma}}$,
it predicts sharp 
conductance peaks split by the magnetic field~\cite{Konig1}.
As stressed in Refs.~\cite{Meir1,Rosch}, we cannot neglect the
decoherence effect, since it could provide the cutoff energy for Kondo
correlations and thus suppresses the Kondo effect. 
In second order perturbation theory, discussed above, 
the decoherence effect appears to be weak,
causing only slight changes of the DOS. 
However, a more thorough analysis reveals that
second order perturbation theory is not enough: 
it does not account for Kondo correlations which also
enhance the decoherence~\cite{Rosch,Paaske2}. 
In order to describe this enhancement, we have to sum up 
infinite order diagrams again.

\begin{figure}[ht]
\includegraphics[width=1.0\columnwidth]{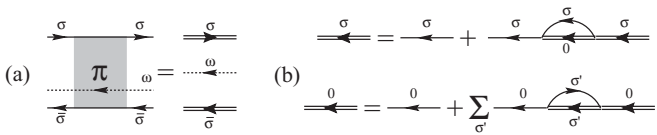}
\caption{ 
(a) The off-diagonal component of the reduced propagator 
in spin space within NCA. 
Double lines on the Keldysh contour are determined by Dyson equations in (b). 
}
\label{fig:NCA}
\end{figure}

A convenient way to include diagrams up to 
the infinite order is provided by the NCA~\cite{Kuramoto,Bickers,Hewson}. 
The naive application of NCA causes a spurious peak at the Fermi energy
(see Sec.~\ref{sec:introduction}), 
but the NCA for the off-diagonal component of the reduced propagator in spin space 
$\pi^{\sigma,\sigma}_{\bar{\sigma},\bar{\sigma}}$
is free from this flaw as shown below.  
The diagram of 
$\pi^{\sigma,\sigma}_{\bar{\sigma},\bar{\sigma}}$ within NCA
is depicted in Fig.~\ref{fig:NCA}(a). 
Double lines are \lq full' propagators of the QD state
and determined by Dyson equations 
generating the infinite numbers of 
\lq non-crossing' diagrams [Fig.~\ref{fig:NCA}(b)]. 
We observe that diagrams in first and second order expansion of 
$\sigma^{\sigma,\sigma}_{\bar{\sigma},\bar{\sigma}}$
[Figs.~\ref{fig:sfgf}(b) and ~\ref{fig:2ndorder}] 
are contained in the NCA diagrams. 
The expression is given by 
\begin{eqnarray}
\pi^{\sigma,\sigma}_{\bar{\sigma},\bar{\sigma}}(\omega)
=
\int \rd \omega_1 \rd \omega_2
\frac{\rho_\sigma(\omega_1) \, 
\rho_{\bar{\sigma}}(\omega_2)}
{\omega+ \ri 0^+-\omega_1+\omega_2 },
\end{eqnarray}
where $\rho_\chi(\omega)$ 
is the spectral density of the
full propagator of the QD state $| \chi \rangle$,
obtained by the following self-consistent equations: 
\begin{eqnarray}
\rho_\sigma(\omega) &=& 
-\frac{1}{\pi} {\rm Im} 
\frac{1}{\omega+\ri 0^+-\epsilon_\sigma-\Sigma_\sigma(\omega)},
\\
\Sigma_\sigma(\omega) &=& 
\int \rd \omega' 
\tilde{\gamma}^-_\sigma(\omega-\omega') 
\rho_0(\omega'),
\\
\rho_0(\omega) 
&=&
-\frac{1}{\pi} {\rm Im} 
\frac{1}{\omega+\ri 0^+-\Sigma_0(\omega)},
\\
\Sigma_0(\omega) &=& 
-\sum_{\sigma'} 
\int \rd \omega' 
\tilde{\gamma}^+_{\sigma'}(\omega'-\omega)^*
\rho_{\sigma'}(\omega').
\end{eqnarray}

\begin{figure}[t]
\includegraphics[width=0.8\columnwidth]{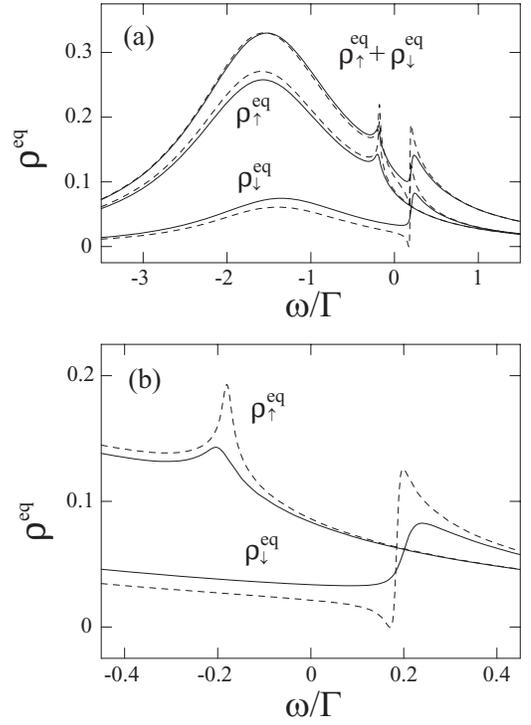}
\caption{ 
(a) The total and spin resolved DOS 
for parallel alignment ($P_L\!=\!P_R\!=\!0.2$) 
within NCA for 
$\pi^{\sigma,\sigma}_{\bar{\sigma},\bar{\sigma}}$ (solid lines). 
The panel (b) is the magnification 
around the Kondo resonance. 
Dotted lines are the results without decoherence effect 
and correspond to lines in Fig.~\ref{fig:linear}(c). 
Parameters are the same as those in Fig.~\ref{fig:linear}. 
}
 \label{fig:decoherence}
\end{figure}

Figures~\ref{fig:decoherence}(a) and (b) 
show the total and spin resolved DOS and their magnifications 
for parallel alignment ($P_L\!=\!P_R\!=\!0.2$) 
within NCA for $\pi^{\sigma,\sigma}_{\bar{\sigma},\bar{\sigma}}$ 
(solid lines). 
By comparing the results without decoherence effect 
[dotted lines correspond to lines in Fig.~\ref{fig:linear}(c)], 
we can see to which extend the Kondo resonance is suppressed by 
the decoherence effect.

Though NCA for 
$\pi^{\sigma,\sigma}_{\bar{\sigma},\bar{\sigma}}$
describes the suppression of the Kondo resonance
and also the spin splitting, careful considerations 
of vertex corrections are necessary~\cite{Paaske2}. 
Moreover additional self-consistency conditions 
require extensive calculations. 
Even if analyzed numerically, this is beyond the scope of the present work, 
the emphasis of which is on spin splitting 
effects in magnetic systems.
For further investigations for the decoherence problem, 
advanced approaches, 
such as the real-time renormalization group 
technique~\cite{Schoeller_Konig} or the generalization of the
perturbative renormalization group to nonequilibrium
states~\cite{Rosch} may be appropriate.

\section{Relation to experiments}
\label{sec:exp}

In this section we will discuss the correspondence between our
calculation and very recent experimental results. 
Pasupathy~{\it et~al.}~\cite{Pasupathy} studied the transport through a
single ${\rm C}_{60}$ molecule attached to ferromagnetic nickel
electrodes in the Kondo regime. It was shown that the Kondo
correlations exist even in the presence of ferromagnetic
leads. The zero-bias anomaly in the nonequilibrium conductance
was split for P alignment of the lead magnetizations, in
agreement with our calculations presented in 
Figs.~\ref{fig:temperature}(a), 7(a), 9(a), and 9(b). 
For AP alignment the
splitting of the zero-bias anomaly was absent in one case - similarly to
Fig.~\ref{fig:nonlinear}(c), and substantially reduced in other cases,
which can be interpreted as an effect of asymmetric coupling
$\Gamma_L \neq \Gamma_R $, similarly to Fig.~\ref{fig:asymmcoupling}(c). The
measurement of the nonequilibrium conductance for the parallel alignment
for several temperatures demonstrate that the splitting of the
Kondo resonance does not depend on temperature, 
in agreement with results presented in Fig.~\ref{fig:temperature}(a).  Also the TMR signal showed very similar properties to those presented in Fig.~\ref{fig:TMR}.

In a different set of experiments Nyg{\aa}rd {\it et~al.}~\cite{Nygard} 
recently observed in the transport
through a carbon nanotube, which acts as a QD,
 attached to normal leads, a zero bias anomaly at low temperature. This
anomaly is split probably due to interaction effects between the carbon
nanotube and a magnetic catalyst particle. These results could be
interpreted within the framework of our model as well.

\section{summary}
\label{sec:sum}

In summary, we have analyzed the nonequilibrium Kondo effect
in a QD coupled to ferromagnetic leads.
We started form a systematic approach --
the real-time diagrammatic technique. We proceeded in an extension of
the `resonant tunneling approximation' by
introducing the self-energy of off-diagonal components of the reduced
propagator in spin space, which was neglected in the original RTA.
The self-energy was calculated to lowest order to describe 
spin-dependent quantum charge fluctuations.
The approximation ensures spin and charge conservation. 
The local magnetization is obtained by 
solving the master equation accounting for quantum fluctuations. 
Our work gives a coherent description of
spin splitting, spin accumulation,
and
Kondo correlations out of equilibrium.
%
%
We calculated the nonlinear differential
conductance for various temperatures. Though the peaks hight of the split
zero-bias anomaly is suppressed with increasing temperatures, the
spin splitting itself is robust against temperature and leaves a clear
structure in the nonlinear differential conductance. Our
approximation provides a reasonable description of the compensation of
the spin splitting by an applied magnetic field.
We also discussed asymmetries in the
zero-bias anomaly in the conductance split by the external
magnetic field for the antiparallel alignment
in term of the spin polarization of the Kondo resonances.
%

We also discussed the effect of the asymmetry in system
parameters. The asymmetry in the coupling strengths does not affect the
spin splitting for the parallel alignment.
 For the antiparallel alignment, the spin
splitting is absent for symmetric set-ups. However, an asymmetry in
coupling strength leads to a spin splitting because the
exchange interaction with the left and the right lead
cannot compensate each other.
The asymmetry is important especially for nonequilibrium experiments, because
it can induce a spin accumulation and affect the weights of the
split Kondo resonance.
We showed that the asymmetry in spin polarization factors of two
leads can induce rather large asymmetries in the peak heights of the
split zero-bias anomaly.

%
%

Though the result looks qualitatively reasonable, 
our lowest-order perturbation theory for 
the self-energy falls
short of a proper description of decoherence effects. 
We showed that the NCA for the 
off-diagonal component of the reduced propagator in spin space 
could account for the decoherence effect which suppresses the 
hight of the Kondo resonances. 
The further analysis of this problem within the real-time approach
remains a challenging problem for the future. 
We hope that the systematic formulation of nonequilibrium Kondo effects in 
magnetic systems presented above encourages further efforts to investigate
these effects in spintronics devices.

\begin{acknowledgments}

We would like to thank
J. Barna\'s,
R. Borda,
P. Bruno,
R. Bulla, 
V. Dugaev,
J. K{\"o}nig,
C. M. Markus,
J. Nyg{\aa}rd,
A. Pasupathy,
D. Ralph,
H. Schoeller,
M. Sindel 
and 
J. von Delft
for valuable discussions and comments.
Y.U. and J.M. were supported by the
DFG-Forschungszentrum 
\lq\lq Centre for Functional Nanostructures" (CFN).
J.M. was also supported by
'Spintronics' RT Network of the EC RTN2-2001-00440, Project PBZ/KBN/044/P03/2001,
and
the Centre of Excellence for Magnetic and Molecular Materials for Future Electronics within the EC Contract G5MA-CT-2002-04049.
S.M. and H.I. were supported 
by CREST, MEXT.KAKENHI(No.~14076204 and No.~16710061), CREST,
NAREGI Nanoscience Project, and NEDO.
\end{acknowledgments}


\begin{thebibliography}{38}
\bibitem{note2}
We introduced vertex functions $\Lambda^\pm$ for compact expressions. 
The relation between 
$\Lambda^\pm$ and 
the function $\phi^{\chi,\sigma}_{\chi,0}$ 
used in the original work~\cite{Konig1} is
$
\phi^{\chi,\sigma}_{\chi,0}(\varepsilou) \! = \! \sum_{s=\pm}
{\Lambda^s}^{\chi,\sigma}_{\chi,0}(\varepsilou)
\gamma^s_\sigma(\varepsilou)
$.
\bibitem{note4}
In addition to the charge conservation, the Kubo-Martin-Schwinger condition 
$G_\sigma^>(\omega) \!=\! G_\sigma^<(\omega) \exp(-\beta \omega)$ 
should be fulfilled to assure vanishing current in equilibrium. However, because the analytic solutions of Eqs.~(\ref{eqn:vertex+}) and (\ref{eqn:vertex-}) 
are difficult to obtain, we only checked that for each numerical result, the current at zero bias voltage is negligibly small.
\end{thebibliography}
\end{document}